\documentclass{iaus}
\usepackage{graphicx}

\title[Access to Synthetic Spectra in the VO] 
{Access to Stellar Population Models in the Virtual Observatory}

\author[Chilingarian]   
{Igor V. Chilingarian$^{1,2,3}$}

\affiliation{$^1$Sternberg Astronomical Institute, Moscow State University,
13 Universitetsky prospect, Moscow, 119992, Russia; email: chil@sai.msu.su\\[\affilskip]
$^2$Centre de Recherche Astronomique de Lyon, Observatoire de Lyon, 9 Av. Charles 
Andr\'e, Saint-Genis Laval, F-69561, France; CNRS, UMR 5574;\\[\affilskip]
$^3$Observatoire de Paris, LERMA, 61 Ave. de l'Observatoire, Paris, 75014, France}

\pubyear{2007}
\volume{241}  
\pagerange{1-2}
\date{30 Jan 2007}
\jname{Proceedings Stellar Populations as Building Blocks of Galaxies}
\editors{A. Vazdekis \& R. Peletier, eds.}
\begin{document}

\maketitle

\begin{abstract}
A great effort is being made by the international Virtual Observatory
community to build tools ready to be used by scientists. Presently,
providing access to theoretical spectra in general, and synthetic spectra of
galaxies in particular, is a matter of current interest in the Virtual
Observatory. Several ways of accessing such spectra are available. We
present two of them for accessing PEGASE.HR evolutionary synthesis models:
HTTP-access to a limited number of parameters using Simple Spectral Access
Protocol (SSAP), and full-featured WEB-service based access using Common
Execution Architecture (CEA).

\end{abstract}

\firstsection 
\section{Introduction}
The main mission of the Virtual Observatory is to increase efficiency of
scientific usage of astronomical data. International Virtual Observatory
Alliance (IVOA\footnote{http://www.ivoa.net/}) is a large international
collaboration of national VO projects. Its main responsibility is to develop
standards within IVOA working groups for the interoperability of archives,
tools, services, and software, and to propose them to the corresponding
division of IAU for further approval and recommendation.

Presently, the Virtual Observatory comprises:
\begin{itemize}
\item Archives and collections of science-ready astronomical data
\item Data access services and analysis tools
\item Resource registries to discover them
\item Client software (interactive tools as well as client API)
\item Set of standards used by all these resources to achieve interoperability
\end{itemize}

Establishing standards for accessing theory data is one of the
cornerstones of present VO development.

\section{Access methods}
Here we present two possible approaches of accessing SSP models. The
PEGASE.HR package (Le Borgne et al. 2004) is taken as an example. Both
methods of accessing PEGASE.HR are implemented in a frame of the VO Paris
project (Simon et al. 2006) by the author.

{\bf TSAP Access}: a subset of the IVOA Simple Spectral Access Protocol
(SSAP, Tody et al. 2006). Parameters are passed in
the HTTP-GET query. Interface is easy to implement. Interactive clients,
such as ESA VOSpec (http://esavo.esa.int/vospec/) are able to build user
interface dynamically, based on the response of the server providing
capabilities (e.g. allowed query parameters). Direct download links to the
FITS files, containing spectral models are provided. Scripting access is
available using standard libraries dealing with HTTP protocol.

{\bf CEA Application Access}: IVOA Common Execution Architecture is a
standard way of accessing generic ``legacy'' applications and codes having CGI
or command-line interfaces. CEA is based on the WebService access interface,
thus parameters of the models and possible additional data are passed as a
part of SOAP message, and no limitation on a number of parameters exists.
Output is saved to the interoperable online storage: VOStore. Interactive
client, ASTROGRID Workbench is available as well as the client library,
ASTROGRID ACR Library, providing scripting level access from various
programming languages, such as Java, C/C++, Perl, Python. Easy integration
into workflows is a valuable advantage of this approach.

\begin{figure}
\begin{tabular}{c|c}
TSAP Access to PEGASE.HR & CEA Access to PEGASE.HR \\
http://vo.obspm.fr/cgi-bin/siap/pegasehr.pl & \\
~~ \includegraphics[width=5.5cm,height=6.0cm]{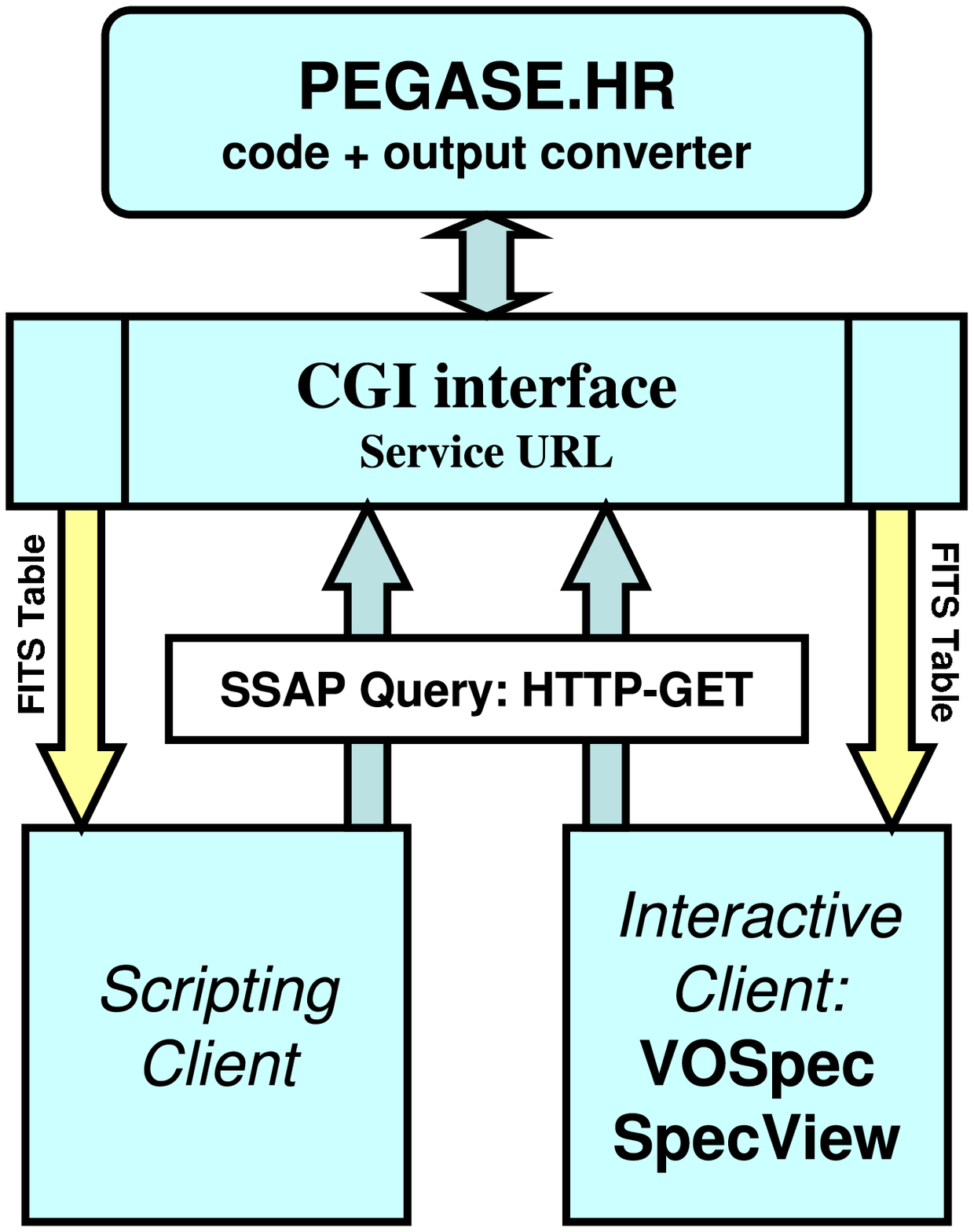} ~~ &
~~ \includegraphics[width=6cm,height=6.0cm]{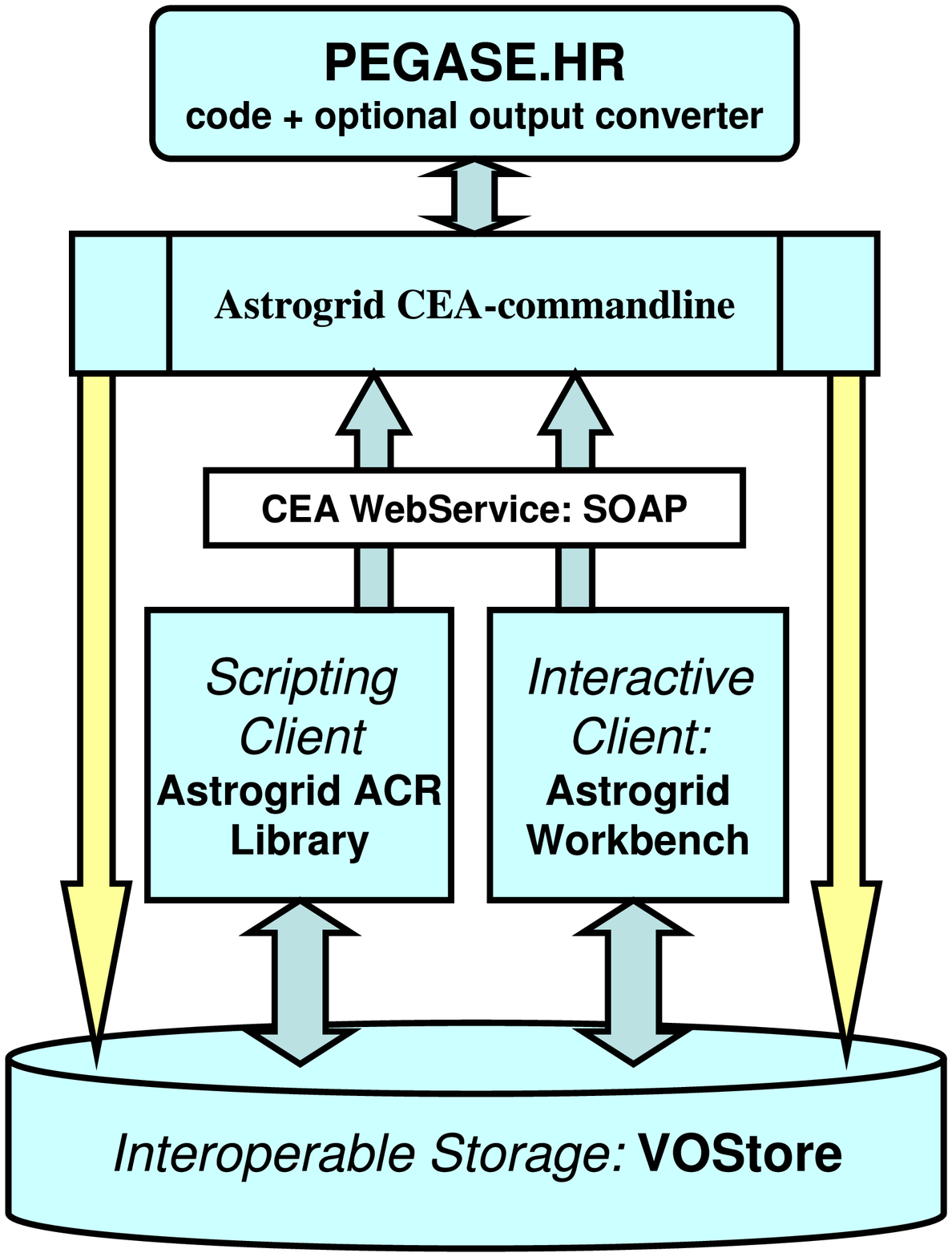} ~~ \\
(a) & (b) \\
\end{tabular}
\caption{Schematic view of two ways of accessing PEGASE.HR simple
stellar population models, available through the VO Paris portal.}
\end{figure}

\section{Conclusion}
CEA Access is more versatile than TSAP, but more expensive in terms of
infrastructure. It is preferable in case of complex models and if used in a
workflow construction

\begin{acknowledgments}
I am very grateful to the financial support, provided by the IAU for
attending this symposium. Special thanks to Florence Durret (Institut
d'Astrophysique de Paris) for additional funding.
\end{acknowledgments}

\end{document}